\documentclass[sigconf]{acmart}
\renewcommand\footnotetextcopyrightpermission[1]{}
\usepackage{booktabs} 
\usepackage{balance} 

\setcopyright{rightsretained}

\copyrightyear{2017} 
\acmYear{2017} 
\setcopyright{acmcopyright}
\acmConference[VAMS'17]{ACM VAMS Workshop}{August 2017}{Como, Italy} 
\acmYear{2017}
\copyrightyear{2017}

\acmPrice{15.00}

\begin{document}

\title{Multiple Stakeholders in Music Recommender Systems}

\author{Himan Abdollahpouri}
\affiliation{%
  \institution{DePaul University}
  \country{USA}
}

\email{habdolla@depaul.edu}
\author{Steve Essinger}
\affiliation{%
  \institution{Pandora Media, Inc.}
  \country{USA} 
}
\email{sessinger@pandora.com}

\begin{abstract}
    Music recommendation services collectively spin billions of songs for millions of listeners on a daily basis. Users can typically listen to a variety of songs tailored to their personal tastes and preferences. Music is not the only type of content encountered in these services, however. Advertisements are generally interspersed throughout the music stream to generate revenue for the business. Additional content may include artist messaging, ticketing, sports, news and weather. In this paper, we discuss issues that arise when multiple content providers are stakeholders in the recommendation process. These stakeholders each have their own objectives and must work in concert to sustain a healthy music recommendation service.
\end{abstract}

%
%


\maketitle

\section{Introduction}
Recommender systems (RS) have been applied successfully in many domains to help users find interesting items and avoid the problem of information overload. Examples of RS include: Amazon, where customers can use recommendations from the system so they can find desirable products easier; Netflix, where the system helps users find interesting movies to watch; Pandora and Spotify, which helps users navigate vast catalogs of content to listen to their favorite artists.
{\let\thefootnote\relax\footnote{Included in the 2017 Workshop on Value-Aware and Multistakeholder Recommendation held in conjunction with RecSys 2017.}}
 
For the most part, the research on RS has been focused on situations where the system has no preference about which product owners the content shown to the user originates from, just as long as the user is happy with the current recommendation.  While this may be acceptable in some applications, there are examples where the recommendations from different stakeholders (e.g. product owners) produce different utilities both to the system and users. For example, if we consider advertisers as stakeholders who want their ads to be shown to the users along with the traditional recommended products, then it is obvious that recommending ads gives different utility to the system than recommending products. The system may make more money from showing a paid ad rather than recommending consumable content. However, from the users' perspective the situation may be different. Users do not really care about ads as much as they want to see their personalized recommendations.

A music recommendation business is an example of managing several different stakeholders that are involved in delivering content. Each stakeholder may have their own incentive to get their content in front of the end user and they may compete or collaborate with each other depending upon the context. In this paper, we discuss how a music recommender service may be viewed as a multi-stakeholder environment where each stakeholder has their own utility function and goal. These stakeholders could include various artists, advertisers or other content providers.

\section{Stakeholders in Music Recommender Systems}
Pandora is one of the world's largest music recommendation companies in terms of active users and songs recommended. The platform consists of a variety of different stakeholders working together to provide users with the best experience, while simultaneously keeping the business alive by generating revenue. Some stakeholders are more user-experience focused so their utility function is closer to what the end user actually wants. Other stakeholders have utility functions which are more revenue-focused and business-oriented. For example, the stakeholder responsible for recommending songs (i.e. music recommender) is focused on delivering the best possible song at the right time to the right end user. Therefore, its objective goal is closely related to what end users want from the system. On the other hand, the stakeholder that is responsible for advertising, \textit{ad service}, is mainly interested in delivering the right ad to the right user at the right time. Generally, users have less interest for ads than songs. Therefore, we can say that the ad service stakeholder's utility is mainly towards more revenue, which is not a primary consideration of the music recommender. 

\begin{figure*}[t]
    \centering
    \includegraphics[width=5in]{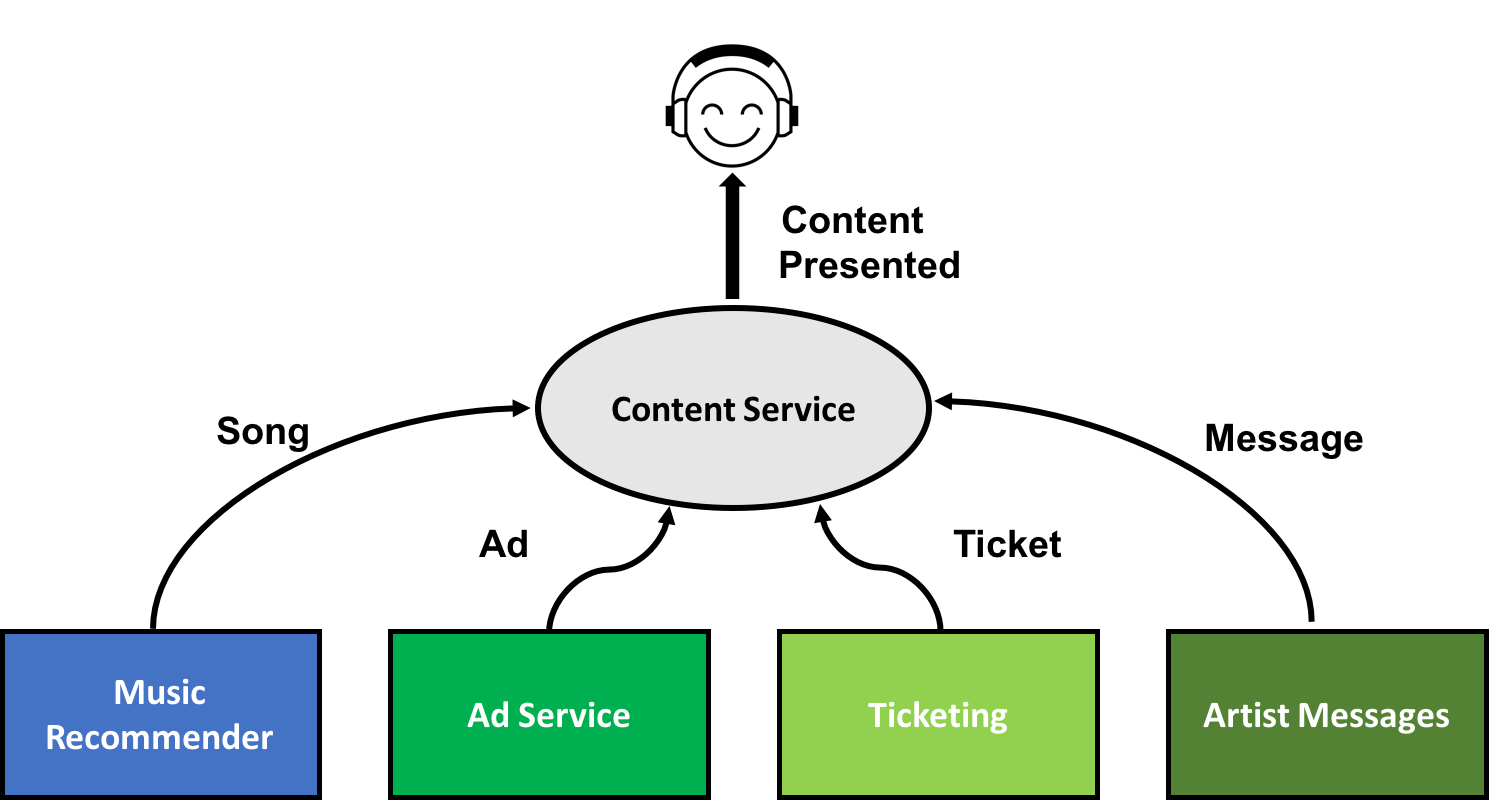}
    \caption{The architecture of a music recommendation platform consisting of multiple stakeholders.}
    \label{fig:ms}
\end{figure*}

\begin{figure*}[t]
    \centering
   \includegraphics[width=5in]{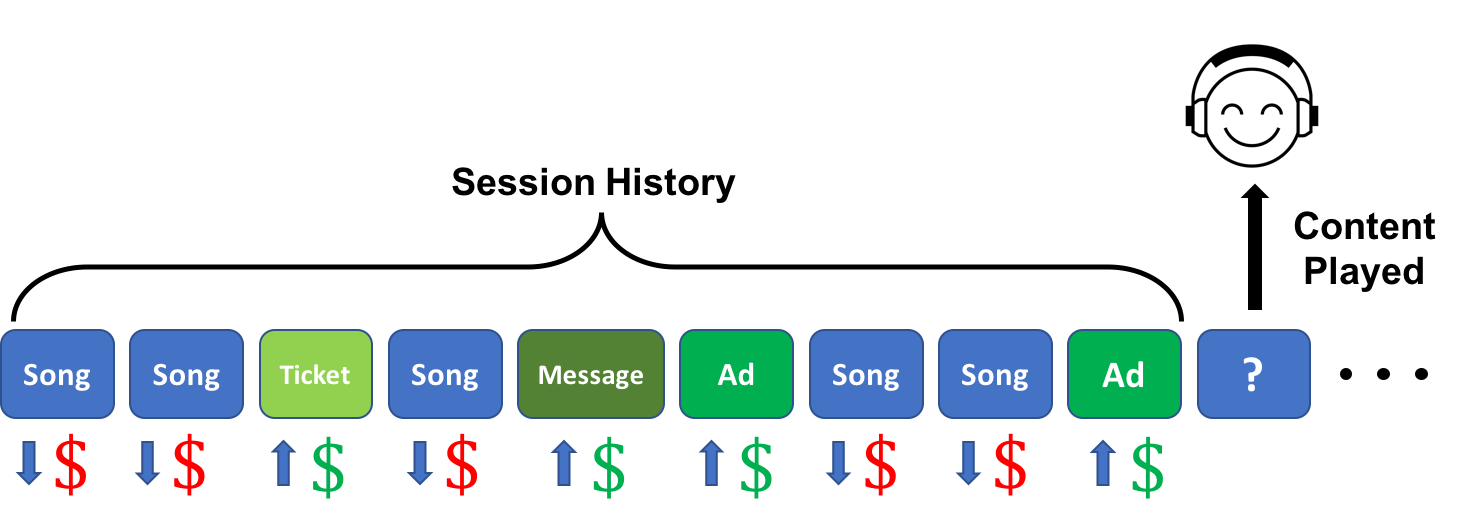}
    \caption{A sequence of various content served to the user. The recommender system should decide what to play for the current slot, indicated by the rightmost box.}
    \label{fig:sequence}
\end{figure*}

Figure ~\ref{fig:ms} shows the architecture of the recommendation platform consisting of different stakeholders. While there may be more fine-grained stakeholders, we only showed four of them for simplicity. For example, the \textit{music recommender} stakeholder is responsible for coming up with the best possible song to play for the user. As we mentioned earlier, this stakeholder is mainly concerned about the needs and interests of the end user. On the other hand, the \textit{ad services} is responsible to find the best time to interrupt the song recommendation to show an ad to the user. In fact, this stakeholder does this calculation based on many different factors such as the contract rules the ad service has with the advertisers, companies financial agenda, etc. There are also a couple of other stakeholders in the platform such as \textit{ticketing} and \textit{artist messages}. The ticketing works similar to the ad services, but instead of ads, there are tickets for which the system tries to find customers. It is important to note that the business rules and utility function for the ticketing stakeholder might be completely different from the ad service even though they both do the similar task of promoting secondary content (i.e. content other than music). The artist messages are recorded audio messages from artists containing information regarding an upcoming concert, album or merchandise. Finally, there is the \textit{content service} that sits on top of the individual stakeholders. This is responsible for managing the different stakeholders and to make the final decision about which content to show to a particular user at a certain time.

In this paper, our focus is mainly on sequential music recommendation where content is recommended one after another. Figure ~\ref{fig:sequence} shows a sequence of different content played to the user in the current session. The sequence contains a variety of content such as songs, ads, artist messages and tickets. Given what has been served to the user thus far, the RS should decide the best content to play for the next slot in the user's session. This slot is shown by the box containing a question mark on the rightmost side of the diagram. All stakeholders try to push their content through the system so that they could be shown to the user. However, since this is a sequential recommendation platform where only a single piece of content can be served at a given time, the content service manager decides which content from one of the stakeholders can occupy this spot.

Certain types of content are more beneficial to the system than others and this is a factor taken into account in the final decision. For example, as shown in  figure ~\ref{fig:sequence}, ads, artist messages and tickets bring money to the system, while songs cost the system money (i.e. there is a royalty per each song played). The balance between the long-term and short-term goals of the system must be considered. Greedy decisions such as playing content based purely on maximizing revenue, while ignoring the end user needs will hurt the system in the long term (i.e. users may not return to the system resulting in fewer users and fewer opportunities to make money via ads). On the contrary, playing only songs will not result in any revenue and the system would not be able to continue service. Therefore, both short-term (make users happy) and long-term goals (run a healthy business) should be considered when deciding what content to play for a user at any given time.

Different artists could also be viewed as stakeholders in terms of what they want from the system. Artists may have differing goals in putting their songs on a music recommendation platform such as Pandora. For example, some artists might be interested in targeting a certain age-group in certain locations and certain times. Perhaps they have a upcoming concert and they want to give more awareness to their targeted audience. In addition, the system might want to promote songs from certain artists to give them a boost in the number of listeners and help them achieve more popularity. However, it is important to take the fairness factors into account in promoting artists so that the system does not over-promote some artists at the price of ignoring others.

\section{Related work}
The concept of multiple stakeholders has been discussed in some prior research. For example, in reciprocal recommendation, a successful recommendation is not the one that is only acceptable by the receiver of the recommendation. In fact, the needs and interests of the other part of the recommendation (i.e. the person who is being recommended) is also important for having a successful match \cite{reciprocal}. Researchers working on multi-objective optimization also investigated the problem of having different objectives and constraints when generating recommendations. These objectives are often different evaluation metrics such as accuracy, popularity bias, novelty and diversity \cite{jugovac2017efficient, biasRecSys2017} or they are some sort of constraint imposing certain limitations on the solution \cite{jambor2010optimizing}. The most explicit definition of having multiple stakeholders in a recommendation setting has been discussed in \cite{umapHimanMS} where authors re-defined the recommender systems as multi-stakeholder environments in which multiple stakeholders involve in different stages of the recommendation generation and each benefit from the recommendation delivery in various ways. For example, in ~\cite{DBLP:conf/um/BurkeAMG16}, the authors discuss a scenario which seeks 'fairness' across multiple product suppliers. Moreover, in \cite{vamsBurkeHiman} authors identify patterns of stakeholder utility that characterize different multi-stakeholder recommendation application, and provide a taxonomy of the different possible systems.  There are many other domains in which multiple stakeholders are involved in the recommendation process. For example, online dating \cite{reciprocal}, educational recommenders  \cite{burke2016educational} and loan recommendation in micro-finance platforms \cite{lee2014fairness} are all instances of recommendations where different stakeholders are involved in receiving or delivering the recommendations.

\balance

\end{document}